\documentclass[preprint,eqsecnum,aps,prd,nofootinbib]{revtex4}
\usepackage[dvips]{graphicx}
\usepackage{graphics, color}
\usepackage{amsmath}
\usepackage{enumerate}
\usepackage{mathrsfs}
\usepackage{amsfonts}

 \newcommand{\be}{\begin{equation}}
\newcommand{\ee}{\end{equation}}

\newcommand{\V}{{\mathcal V}}

\newcommand{\op}{{\mathcal O}}

\newcommand{\halpha}{{\hat \alpha}}
\newcommand{\hbeta}{{\hat \beta}}
\newcommand{\hs}{{\hat s}}

\newcommand{\heps}{{\epsilon}}

\usepackage{hyperref}

\begin{document}

\title{Stability in Einstein-Scalar Gravity with a Logarithmic Branch}

\author{Aaron J.~Amsel}
\email{aamsel@asu.edu}
\affiliation{Department of Physics and Beyond Center for Fundamental Concepts in Science, \\ Arizona State University, \\Tempe, AZ 85287}

\author{Matthew M.~Roberts}
\email{matthew.roberts@nyu.edu}
\affiliation{Center for Cosmology and Particle Physics, Department of Physics, New York University, 4 Washington Place, New York, NY 10003}

\begin{abstract}
We investigate the non-perturbative stability of asymptotically anti-de Sitter
gravity coupled to tachyonic scalar fields with mass saturating the Breitenlohner-Freedman bound.
Such ``designer gravity'' theories admit a large class of boundary conditions at asymptotic infinity.
At this mass, the asymptotic behavior of the scalar field develops a logarithmic branch, and previous attempts at proving a minimum energy theorem failed due to a large radius divergence in the spinor charge.   In this paper, we finally resolve this issue and derive a lower bound on the conserved energy. Just as for masses slightly above the BF bound, a given scalar potential can admit two possible branches of the corresponding superpotential, one analytic and one non-analytic.  The key point again is that existence of the non-analytic branch is necessary for the energy bound to hold. We discuss several AdS/CFT applications of this result, including the use of double-trace deformations to induce spontaneous symmetry breaking.
\end{abstract}

\maketitle
\tableofcontents

\section{Introduction}

The bulk side of the AdS/CFT correspondence \cite{Maldacena:1997re, Witten:1998qj, Gubser:1998bc} consists of gravity coupled to various matter fields.  In particular, supergravity compactifications relevant to AdS/CFT \cite{Gunaydin:1984fk, Kim:1985ez, Gunaydin:1984wc} often contain tachyonic scalar fields with masses at or slightly above the Breitenlohner-Freedman (BF) bound \cite{BF}.  In some cases, the bulk theory can be consistently truncated so that the matter content is just scalar fields \cite{Duff:1999gh, Freedman:1999gk}.

Such theories of AdS$_{d+1}$ gravity coupled to scalar fields near the BF bound (sometimes called ``designer gravity'' \cite{HH2004}) are known to admit a large class of boundary conditions, which can be defined in terms of an arbitrary function $W$.  The scalar fields have slower fall-off than allowed by the standard asymptotically AdS boundary conditions of \cite{Henneaux:1985tv}, but nevertheless, the conserved charges have been shown to be finite and well-defined once back-reaction effects are taken into account \cite{Henneaux04,Henneaux:2006hk,Amsel:2006uf}. This paper is concerned with the conditions under which the total conserved energy is bounded from below (for other interesting applications, see e.g., \cite{Hertog:2003zs,Hertog:2003xg,HM2004,Hertog:2004gz,Hertog:2004rz,Hertog:2004jx,Hertog:2005hu,Hertog:2006rr,Craps:2007ch,Battarra:2011nt}).

The derivation of the energy bound proceeds by following a Witten-Nester style argument using a spinor charge  \cite{Witten:1981mf,Nester:1982tr}.
For the standard or ``Dirichlet'' scalar boundary conditions (i.e., when the leading, slower fall-off term in the asymptotic expansion is turned off), it was proven several decades ago that the energy is positive if the scalar potential is generated by a superpotential \cite{Gibbons:1983aq,Boucher:1984yx,Townsend:1984iu}.  More recently, this proof was extended to the more general slow fall-off designer gravity boundary conditions in  \cite{Amsel:2006uf,Amsel:2007im} (based on \cite{Hollands:2005wt,Hertog2005}), where it was shown that the theory is stable if $W$ is bounded from below \emph{and} the scalar potential admits a certain type of superpotential.   This minimum energy theorem was then further strengthened to allow stability even in some cases when $W$ is unbounded from below, so long as the full effective potential $\V$ (defined below) has a global minimum \cite{Faulkner:2010fh}.  This result finally proved a conjecture about stable ground states in designer gravity that was originally given in \cite{HH2004}.

However, the stability conjecture of \cite{HH2004} was never proven in the special case where the BF mass bound is saturated.  This case requires separate treatment, as the asymptotic behavior of the scalar field develops a logarithmic branch\footnote{In fact, there are additional special values of the mass for which such logarithmic modes may appear in the asymptotic expansion of the scalar field \cite{Henneaux:2006hk,Amsel:2006uf}.  These cases require a separate treatment as well.}. While the theory is known to be stable if the logarithmic branch is turned off, previous attempts at proving a minimum energy theorem for more general boundary conditions failed due to a logarithmic large radius divergence in the spinor charge \cite{Amsel:2006uf} .  In this paper, we resolve this issue and derive a minimum energy bound, which agrees with the conjecture of \cite{HH2004}.  Once again, the main subtlety involves the existence of a suitable superpotential for a given scalar potential.

It is a general principle of AdS/CFT that deformations of the CFT correspond to modifications of the AdS boundary conditions.  For designer gravity theories with a field theory dual, the boundary conditions given by the function $W$ are related to the addition of a multi-trace potential term $\int d^d x~W(\op)$ to the CFT action \cite{witten, bss,Sever:2002fk}, where $\op$ is the operator dual to the bulk scalar.  The effective potential $\V(\op)$ is simply the effective lagrangian of the CFT restricted to constant values of $\op$ (and all other fields and currents turned off).  (See, for example, \cite{Papadimitriou:2007sj,Vecchi:2010dd} for discussion of multi-trace deformations and stability from the dual field theory perspective). The interesting point about the result of \cite{Faulkner:2010fh} is that by adding an unbounded potential term to the CFT, it is possible to destabilize the AdS vacuum but still have a stable ground state, leading to spontaneous symmetry breaking.  In \cite{Faulkner:2010gj}, relevant double-trace deformations were used to create a novel type of holographic superconductor, which, in contrast to previous constructions \cite{Hartnoll:2008vx}, can exist without a net charge density (see also \cite{Kiritsis:2011zq,Iqbal:2011aj}).

When the bulk scalar saturates the BF bound, the dual operator has dimension $d/2$ in both the standard Dirichlet and alternate Neumann theories, and therefore a double trace term $\int d^dx~\op^2$ is classically marginal.  As first pointed out in \cite{witten}, double-trace deformations of the Dirichlet theory lead to a logarithmic running of the coupling.  The deformation can be asymptotically free with an infrared Landau pole, or marginally irrelevant with a UV Landau pole, depending on the sign of the coupling.

In the alternate Neumann theory, the double-trace coupling is marginally irrelevant, in the sense that it diverges logarithmically in the UV.  At zero temperature  and with planar symmetry, we will show below that the AdS vacuum is always unstable to the true ground state with $\langle\op\rangle \neq 0$, independent of the double-trace coupling.  At sufficiently high  temperature, however, the system returns to the symmetry-preserving state and we demonstrate the existence of a superconducting phase transition at the critical temperature.  The energy in the  asymptotically Poincar\'e AdS case is always bounded from below by the energy of the zero-temperature ordered state, which corresponds to the global minimum of the effective potential, verifying our expectation from previous designer gravity work.  To summarize the main result of this paper, we prove the energy bound in the Neumann theory is given explicitly by
\begin{equation}\label{boundstatement}
E \geq \mathrm{Vol}(S^{d-1}) \,\mathrm{inf}\left[W(\alpha)
+C \alpha^2 +\frac{1}{2d} \alpha^2 \log\alpha^2 \right] \,,
\end{equation}
where $\alpha = \langle \op \rangle$ is the coefficient of the logarithmic term in the asymptotic expansion of the scalar field, and $C$ is a constant which we will specify later.  The term in square brackets turns out to be the (zero-temperature) effective potential $\V(\langle\op\rangle)$ in the large $\alpha$ limit (and is exactly $\V$ in the planar case).

This paper is organized as follows.  In section \ref{sec:designergravityreview}, we give a more detailed introduction to designer gravity and review previous work on minimum energy theorems in these theories.  In section  \ref{sec:superpotential}, we find a new branch of superpotential solutions in the case where the BF bound is saturated.  We show that this superpotential (if it exists globally) cures the divergent spinor charge encountered in \cite{Amsel:2006uf} and we derive a lower bound on the energy.  Section  \ref{sec:fakesugra} focuses on planar, boost-invariant solutions, which turn out to saturate the bound.  We argue in section  \ref{sec:solitons} that these ``fake supergravity'' solutions correspond to a certain limit of spherical solitons, which leads to a proof of the stability conjecture of  \cite{HH2004}.  Several AdS/CFT applications of this result are investigated in section  \ref{sec:poincare}, including the generalization to finite temperature.  These results refer to deformations of the Neumann theory, so in section \ref{sec:dirichlet} we briefly examine some of the corresponding issues for deformations of the Dirichlet theory.  We close with a discussion of our results in section \ref{sec:discussion}.


\section{Designer Gravity Review}
\label{sec:designergravityreview}

In this section, we briefly review the important features of designer gravity theories.  We focus in particular on the stability conjecture of  \cite{HH2004} and we describe previous efforts to prove this conjecture.

We consider asymptotically AdS$_{d+1}$ gravity ($d \geq 3$) coupled to a tachyonic scalar field with action
\begin{equation}
\label{theory}
S = \frac{1}{2} \int \, d^{d+1} x \sqrt{ - g} \, [R -
(\nabla \phi)^2 - 2V(\phi)]  \, ,
\end{equation}
where we have set $8\pi G = 1$.  Near $\phi = 0$, we assume that the scalar potential $V(\phi)$
takes the form
\begin{equation}
\label{potential}
V(\phi) = -\frac{d(d-1)}{2\ell_{AdS}^2}+\frac{1}{2} m^{2} \phi^{2} + \ldots \,,
\end{equation}
where $\ell_{AdS}$ is the AdS radius.   It will be convenient to work in units where
$\ell_{AdS} =1$.
Unless stated otherwise, we consider only even potentials for simplicity, though our results easily generalize to non-even potentials\footnote{{The generalization is given by constructing the critical superpotential (as described below) for both  $\phi >0$ and $\phi <0$, which will provide two different values of the constant $C$ in (\ref{boundstatement}). The bound is then simply (\ref{boundstatement}) with $C=\max(C_>,C_<).$}}.
 In designer gravity theories,
we restrict to scalar masses $m^2 < 0$ in the range
\begin{equation}
\label{range}
m^2_{BF} \le m^2 < m^2_{BF} +1,
\end{equation}
where the Breitenlohner-Freedman bound for perturbative stability \cite{BF} is
\begin{equation}
m^2 \geq m^2_{BF} =-\frac{d^{\,2}}{4} \,.
\end{equation}

We are interested in metrics which asymptotically approach \cite{Hollands:2005wt,Hertog2005,Amsel:2006uf} the metric of exact AdS
spacetime in global coordinates,
\begin{equation}
\label{pureads}
ds^{2} = - \left(1 + r^2 \right)\, dt^2 +
\frac{dr^2}{1 + r^2} + r^2  d \Omega^2_{d-1}\, .
\end{equation}
Here $d \Omega^2_{d-1}$ is the metric on the sphere $S^{d-1}$.
For most masses in the range \eqref{range},
the scalar field behaves near the AdS boundary  ($r\to\infty$) as
\begin{equation}
\label{phi}
\phi = \frac{\alpha}{r^{\lambda_{-}}} + \frac{\beta}{r^{\lambda_{+}}}  + \dots \,,
\end{equation}
where
\begin{equation}
\label{roots}
\lambda_{\pm} = \frac{d \pm \sqrt{d^{\,2}+ 4 m^2}}{2} \,,
\end{equation}
and the coefficients $\alpha, \beta$ do not depend on the radial coordinate $r$.
For $m^2 = m^{2}_{BF}$, the roots \eqref{roots} are degenerate and the solution has the asymptotic behavior\footnote{See \cite{Henneaux:2006hk,Amsel:2006uf} for discussion of additional cases where logarithmic branches may arise.  In general, this can occur when $\lambda_+/\lambda_- = n$, where $n$ is an integer.  The present work is concerned with the case $n=1$.  See also \cite{resonant}.}
\begin{equation}
\label{phibf}
\phi = \frac{\alpha \,\log r}{r^{d/2}} + \frac{\beta}{r^{d/2}} + \dots\,.
\end{equation}
Note that in global AdS we use the radius of the boundary $S^{d-1}$ to define the scale of the logarithm. This means that one should interpret $\log r = \log(r/R_{S^{d-1}}),$ and $R_{S^{d-1}}=\ell_{AdS}=1$ in our units.

In the mass range \eqref{range}, both the $\alpha,\beta$ modes are normalizeable, but in order to have well-defined evolution we must impose a
boundary condition at the AdS boundary.  For example, the standard Dirichlet boundary condition is to fix $\alpha = 0$.  Alternatively, one could choose the Neumann boundary condition $\beta = 0$.  More generally, it is sufficient to fix a functional relation between $\alpha$ and $\beta$, which we express as
\begin{equation}
\label{dW}
\beta \equiv \frac{dW}{d\alpha} \, ,
\end{equation}
for some arbitrary smooth function $W(\alpha)$.
Note that a general boundary condition $W$ will break the asymptotic AdS symmetry, but conformal invariance is preserved by the choice
\begin{eqnarray}
\label{cibc}
W(\alpha) &=& k |\alpha|^{d/\lambda_-} \,, \qquad \qquad \qquad m^2 \neq m^2_{BF} \\
\label{bfbc}
W(\alpha) &=& k \alpha^2-\frac{1}{d} \alpha^2 \log |\alpha|\,, \qquad m^2 =m^2_{BF}\label{BFWa}
\end{eqnarray}
for some arbitrary constant $k$.  It is worth noting that for $m^2 \neq m^2_{BF}$ the Neumann theory $W(\alpha)=0$ preserves the conformal symmetry.   However, this is not true for $m^2 = m^2_{BF}$, since the Neumann boundary condition does not include the logarithmic term in (\ref{BFWa}). (Dirichlet boundary conditions $\alpha=0$ of course always preserve the conformal symmetry.)

Solitons are nonsingular, static, spherically symmetric solutions of the bulk gravity theory.
We expect the minimum energy ground state of a designer gravity theory to be given
by one of these solitons \cite{HH2004, Hertog2005}.  For every choice of $\phi$ at the
 origin, the solutions to the equations of motion behave as in \eqref{phi}
 or \eqref{phibf} for some (constant) values of $\alpha, \beta$.  By scanning different values for $\phi(0)$, we map out a curve in the $\alpha, \beta$ plane\footnote{For certain scalar potentials, this curve may not be
 single-valued, so it does not define a function $\beta_0(\alpha)$.  For example, the known supergravity truncations containing scalars at the BF bound (see e.g., \cite{HM2004,Craps:2007ch}) appear to exhibit this behavior.  We will generally not consider such cases in this work, though we do make some further comments in section \ref{sec:discussion}.}, which we call $\beta_0(\alpha)$.  The solitons consistent with our boundary conditions are then given by the intersection points, $\beta_0(\alpha) = W'(\alpha)$.  Let us now define
\begin{equation}
\label{W0def}
W_0(\alpha) = -\int^\alpha_0 \beta_0(\tilde \alpha) d\tilde \alpha \,,
\end{equation}
and
\begin{equation}
\label{effV}
\V(\alpha) = W(\alpha) +W_0(\alpha) \,.
\end{equation}
It was shown in \cite{HH2004} that extrema of $\V$ (denoted $\alpha =
\alpha_*$) correspond to solitons satisfying our boundary conditions,
and further that the value of $\V(\alpha_*)$ gives the total energy  of the
soliton (up to overall volume normalization).

The above statements translate simply to the field theory side. The bulk scalar is dual to an operator $\op$ of conformal dimension $\Delta = \lambda_-$ (which becomes $\Delta = d/2$ when $m^2 = m^2_{BF}$).  Our boundary conditions \eqref{dW} correspond to a deformation of the Neumann theory\footnote{An equally valid boundary condition would be to choose $\alpha = \alpha(\beta)$, which would correspond to a deformation of the Dirichlet theory $\alpha = 0, \Delta = \lambda_+$.  This case will be addressed further in section \ref{sec:dirichlet}, so for now we restrict our discussion to deformations of the Neumann theory.}   by adding a term to the action
\be
\Delta S_{CFT} = - \int d^d x ~ W(\op).
\ee
The function $\V$ is simply the effective potential for the operator, which is minus the effective action restricted to constant field configurations (see e.g., \cite{Papadimitriou:2007sj}),
\be
\V(\alpha) = - \Gamma[\op(x^\mu)=\alpha] \,.
\ee
Every soliton corresponds to an extremum of $\V$ with $\langle \op
\rangle = \alpha_*$, and the energy of the state is simply $\int d^dx ~ \V(\alpha_*)$.
Based on this interpretation, it was conjectured in \cite{HH2004} that the theory would be stable if $\V$ admits a global minimum.  We now briefly review previous work on proving this conjecture.

\subsection{Stability for General Scalar Mass}

We first assume $m^2 \neq m^{2}_{BF}$.  The minimum energy bound is derived
following a Witten-Nester style proof \cite{Witten:1981mf,Nester:1982tr,Gibbons:1983aq,Boucher:1984yx,Townsend:1984iu}, which makes use of the spinor charge
\begin{equation}
\label{charge}
Q = \int_{C} \ast {\bf B} \,, \quad B_{cd} = \bar{\psi} \gamma_{[c}\gamma_d\gamma_{e]} \widehat{\nabla}^e \psi + \textrm{h.c.}\, ,
\end{equation}
where $C =\partial \Sigma$ is a surface at spatial infinity that bounds a
spacelike surface $\Sigma$.  The covariant derivative is
\begin{equation}
\widehat{\nabla}_a \psi = \nabla_a \psi +
\frac{P(\phi)}{\sqrt{2(d-1)}} \gamma_a \psi \,,
\end{equation}
where the ``Witten spinor'' $\psi$ is required to satisfy a spatial Dirac equation $\gamma^i \widehat{\nabla}_i \psi = 0$ and to asymptotically approach a Killing spinor of exact AdS (see e.g., \cite{Amsel:2007im}). Using standard manipulations, it can be shown that $Q \geq 0$ if the
``superpotential'' $P$ satisfies
\begin{equation}
 \label{vtop}
 V(\phi) = (d-1) \left(\frac{dP}{d\phi}\right)^2 -d P^2 \, .
\end{equation}
Using the perturbative solutions for small $\phi$,\footnote{ Here there is as an additional assumption that $P'(0) = 0$, as otherwise there would be additional divergent terms in $Q$.}
\begin{equation}
\label{series}
P_{\pm}(\phi) = \sqrt{\frac{d-1}{2}} +\frac{\lambda_\pm}{2\sqrt{2(d-1)}} \, \phi^2 + O(\phi^4) \, ,
\end{equation}
it was shown in \cite{Amsel:2007im} that
\begin{equation}
\label{qew}
Q_\pm = E -  \oint \left[(\lambda_+ -\lambda_-)W(\alpha)- (\lambda_{\pm} - \lambda_-)\alpha \beta \right]
 + \frac{1}{2} \lim_{r \to \infty} (\lambda_{\pm}-\lambda_-)
\, r^{d-2\lambda_-} \oint \alpha^2 \,,
\end{equation}
where the integrals are over the unit sphere $S^{d-1}$.  Here $E$ is
the total conserved energy, whose explicit form can be found in
\cite{Amsel:2006uf}. If we use the $P_+$ superpotential to construct
the spinor charge, the last term in \eqref{qew} diverges, and we do not obtain a bound on the energy (except of course in the Dirichlet theory, $\alpha=0$). If
instead we use the $P_-$ superpotential, the divergent terms cancel
and we obtain
\begin{equation}
\label{bound}
E \geq (\lambda_+ - \lambda_-)\oint W \geq \textrm{Vol}(S^{d-1})  (\lambda_+ - \lambda_-)
\inf W \,.
\end{equation}
So the energy is bounded from below if $W$ has a global minimum \emph{and}
the scalar potential can be generated by a real $P_-$-type
superpotential that exists for all $\phi$.
Note, however, that this result is slightly weaker than the original conjecture of
\cite{HH2004}.
Furthermore, the linearized analysis of
\cite{Ishibashi:2004wx} suggested that designer gravity theories could be stable
even in some cases where $W$ is not bounded from below.

The  result \eqref{bound} was eventually generalized in \cite{Faulkner:2010fh} by noting
that the $P_-$ branch can be extended to a one-parameter family of
solutions \cite{Papadimitriou:2006dr, Papadimitriou:2007sj}
\begin{equation}
\label{series2}
P_s(\phi) = \sqrt{\frac{d-1}{2}} +\frac{\lambda_-}{2\sqrt{2(d-1)}} \, \phi^2 -\frac{\lambda_- (\lambda_+ - \lambda_-)}{d \sqrt{2(d-1)}} \,s |\phi|^{d/\lambda_-}+\ldots
\end{equation}
for any $s$.  Note that the $P_+$ solution is isolated from the family $P_s$ and does not have an analogous generalization.
Once again,  $P_s$ is required to be real and exist globally, and generally this holds up to some critical value $s_c >0$.  Hence, \eqref{bound} becomes
\begin{equation}
\label{sbound}
E \geq (\lambda_+ - \lambda_-)\oint \left[W+\frac{\lambda_-}{d} \, s_c |\alpha|^{d/\lambda_-}\right] \,.
\end{equation}
Using scaling arguments, one can show \cite{Faulkner:2010fh} that for large $\alpha$ we have
\begin{equation}
W_0(\alpha) = \frac{\lambda_-}{d} \,s_c |\alpha|^{d/\lambda_-} \,,
\end{equation}
so that if $\V = W+W_0$ is bounded, the right hand side of \eqref{sbound} is bounded. This proved the conjecture stated above.  The result confirms that it
is possible for the
theory to be stable even in some cases where $W$ is not bounded from
below.


\subsection{Stability at the BF Bound}
\label{reviewBF}

We now review previous results for $m^2 = m^{2}_{BF}$.  In this case, the superpotential branches \eqref{series} apparently degenerate to a single solution
\begin{equation}
\label{Pplus}
P_{+}(\phi) = \sqrt{\frac{d-1}{2}} +\frac{d}{4\sqrt{2(d-1)}} \, \phi^2 + O(\phi^4) \, .
\end{equation}
(The reason for using the $P_+$ notation here will be clear shortly.)
Repeating the calculation of the spinor charge produced \cite{Amsel:2006uf}
\begin{equation}
\label{qtoh}
Q_+ = E  - \lim_{r\to\infty}\oint\left[W -\alpha \beta -\frac{\alpha^2}{2} \log r \right] \geq 0 \,.
\end{equation}
Once again, the explicit expression for the (finite) conserved energy is
given in \cite{Amsel:2006uf}.  When the logarithmic mode is turned
off, the energy is positive \cite{Hollands:2005wt}.
However, for $\alpha \neq 0$,
the expression for the spinor charge diverges in $r$,
so this result did not yield an energy bound.
Note that this is exactly what happened for the $P_+$ branch of superpotentials
when $m^2 \neq m^2_{BF}$.  Based on this, we might expect that at the
BF bound, there also exists a second branch\footnote{The existence
of this second branch of superpotential solutions for $m^2 = m^2_{BF}$ was previously noted in
\cite{Papadimitriou:2007sj}, though not all relevant terms in the
small $\phi$ expansion were given.} of solutions (analogous to
\eqref{series2}) for which the spinor charge would be finite in $r$.
We will now show that this is indeed the case.


\section{The Superpotential}
\label{sec:superpotential}
In this section, we find a second branch of superpotential solutions at the BF bound, which is analogous to the $P_-$ branch discussed above.  We show that constructing the spinor charge using this type of superpotential does in fact lead to a minimum energy theorem.

We wish to find a new solution to \eqref{vtop} which will cancel the
divergent term that appears in the spinor charge \eqref{qtoh}.  It is
straightforward to check that $\alpha^2 \log{r} \sim \phi^2/\log
\phi$ for large $r$.  With this motivation,
we consider general superpotentials of the form
\begin{equation}
P(\phi) = \sqrt{\frac{d-1}{2}}+ p_0 \phi^2+ p_1 \, \frac{\phi^2}{\log |\phi|} +\ldots
\end{equation}
Substituting this expansion for $P$ and \eqref{potential} for $V$ into
\eqref{vtop}, we find that the only ($p_1 \neq 0$) solution which
simultaneously cancels the $O(\phi^2)$ and $O(\frac{\phi^2}{\log \phi})$ terms is
\begin{equation}
\label{p0}
m^2 = -\frac{d^2}{4} \,, \qquad p_0 = \frac{d}{4 \sqrt{2(d-1)}} \,,
\end{equation}
i.e., the BF bound is saturated.  Continuing to higher order terms in the expansion, we have
\begin{eqnarray}
\label{bfsp}
P_s(\phi) &=& \sqrt{\frac{d-1}{2}}+p_0 \phi^2+ p_0 \,\frac{\phi^2}{\log |\phi|}
+p_0 \, \frac{\phi^2 \log|\log |\phi||}{(\log |\phi|)^2}+s \, \frac{\phi^2}{(\log |\phi|)^2}\nonumber \\
&& \qquad+ \,O\left(\frac{\phi^2 (\log|\log |\phi||)^2}{(\log |\phi|)^3}\right) \,.
\end{eqnarray}
The parameter $s$ is not fixed by the relation \eqref{vtop}, and the coefficients of all higher order terms are given in terms of $s$.

We can now repeat the calculation of the spinor charge using the new small $\phi$ solution \eqref{bfsp}.
The $O\left(\frac{\phi^2 (\log|\log \phi|)^2}{(\log \phi)^3}\right)$ and higher order terms fall off fast enough at infinity that they do not contribute to the spinor charge.  Assuming global existence of the superpotential (see below), the energy bound becomes\footnote{This result still holds if we relax the assumption that $V(\phi)$ is even.  In particular, including a $O(\phi^3)$ term in \eqref{potential} would not affect \eqref{bfboundfinal}.}
\begin{equation}
\label{bfboundfinal}
E \geq \oint\left[W -\frac{4 s\sqrt{2(d-1)} + d \log(d/2)}{d^2}
\alpha^2 +\frac{1}{2d} \alpha^2 \log\alpha^2 \right] \,.
\end{equation}
Note that the original $\log r$ divergence in \eqref{qtoh} is canceled
by the $O\left(\frac{\phi^2}{\log \phi}\right)$ term in the
superpotential.  This term also leads to a new term diverging as
$\log(\log r)$, but this new divergence is conveniently canceled by
the $O\left(\frac{\phi^2 \log|\log \phi|}{(\log \phi)^2}\right)$ term.
The contribution from the $O\left(\frac{\phi^2}{(\log \phi)^2}\right)$
term is finite.  In appendix~\ref{limit}, we
derive this result again using a different method in which we take the limit $m^2 \to
m^2_{BF}$.

Recall that for $m^2 \neq m^2_{BF}$, the minimum energy result failed
when using the $P_+$ type superpotential \eqref{series}, due to a large $r$
divergence in the spinor charge.  Further, the $P_+$ branch is
isolated from the one-parameter family of solutions
\eqref{series2}, which does lead to a minimum energy theorem.
We now see that the situation when the BF bound is saturated is quite
similar.  The original solution \eqref{Pplus} is isolated from the new
one-parameter family of
solutions superpotential \eqref{bfsp} and does not yield a bound on the energy.  Hence, one may think of \eqref{Pplus} as ``$P_+$ type."  Meanwhile, the solution \eqref{bfsp} should be considered ``$P_-$ type," which is consistent with the fact that this superpotential does produce a minimum energy theorem.


\subsection{The Critical Superpotential}

\begin{figure}
\begin{picture}(0,0)
\put(-175,-100){$P_s'(\phi)$} \put(0,-205){$\phi$}
\end{picture}
\begin{center}
\includegraphics[width=4in]{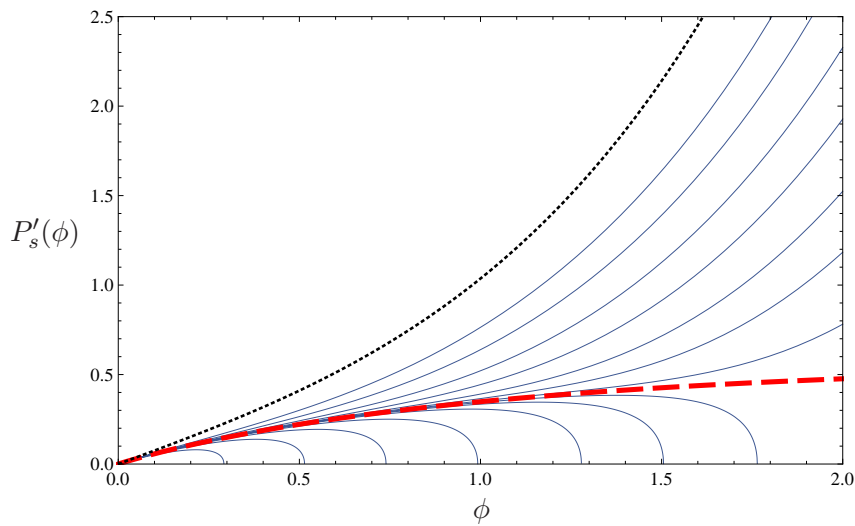}
\end{center}
\caption{Here we plot $P_s'(\phi)$ for the simple potential
\eqref{simpleV} and for various values of $s$.  For
$s$ less than the critical value, $P_s'$ vanishes at some $\phi$ and
so $P_s$ does not exist globally.  The red dashed curve is the critical
superpotential and corresponds to $s = 0.35$.  The black dotted
curve corresponds to a solution whose small $\phi$ behavior is given
by $P_+$ in \eqref{Pplus}. } \label{SPplot}
\end{figure}
The solution \eqref{bfsp} is always valid perturbatively near
$\phi = 0$.  However, the derivation of the energy bound requires the
existence of a real superpotential for all $\phi$.  In general, the
full solution to \eqref{vtop} can only be found numerically.  For this
it is convenient to rewrite
\begin{equation}
\frac{dP}{d\phi} = \pm \frac{1}{\sqrt{d-1}} \, \sqrt{V+d P^2} \,.
\end{equation}
This can be solved by integrating out from $\phi = 0$ and matching to
\eqref{bfsp} for small $\phi$.  The solution fails to exist if the
quantity under the square root becomes negative.  Similar to
\cite{Faulkner:2010fh}, we expect $P_s$ to exist globally above some
critical value $s_c$. In all cases studied, this is indeed the behavior we find. Therefore, the strongest energy bound is (\ref{bfboundfinal}) with $s=s_c$. Note that unlike the case away from the BF bound, the sign of $s_c$ is not important to the stability of the Neumann theory,  as the $\alpha^2$ term is dominated by the positive $\alpha^2 \log \alpha$ term at large $\alpha$.

For example, consider the simple potential
\begin{equation}
\label{simpleV}
\bar V(\phi) = -3+\frac{1}{2} m_{BF}^2\phi^2\,, \qquad d = 3 \,.
\end{equation}
We wish to determine whether or not $P_s(\phi)$ exists globally as we vary the parameter $s$.
For $s<0.35$, we find that there is some $\phi$ at which $P_s'$ becomes imaginary, so a global real solution does not exist.
For $s \geq 0.35$, we find that $P_s(\phi)$ exists globally.
Numerical solutions for various values of $s$ are plotted in Figure~\ref{SPplot}.
Now, the strictest possible energy bound \eqref{bfboundfinal}  will be
given by the \emph{minimum} value of $s$ for which the
superpotential exists globally.   Hence, the critical $s$ which
appears in the energy bound is $s_c = 0.35$.  The behavior is similar in all other examples that we have tested, though of course the precise value of $s_c$ depends on the choice of $V(\phi)$.


\section{Fake Supergravity}\label{sec:fakesugra}
\label{fakesugra}

In this section, we analyze a class of planar domain walls in Einstein-scalar gravity.  As explained below, these solutions turn out to be related to the static, spherical solitons referred to in the stability conjecture of \cite{HH2004}.

Following \cite{Faulkner:2010fh}, we consider boost-invariant planar solutions of the form
\begin{equation}
ds^2 = r^2(-dt^2 +d\vec{x}^2)+\frac{dr^2}{g(r)} \,, \qquad \phi = \phi(r) \,.
\end{equation}
When the potential can be derived from a superpotential, it follows
\cite{DeWolfe:1999cp,Freedman:2003ax} that
\begin{equation}
\label{eomfake}
\frac{d\phi}{dr} = -\frac{(d-1)}{r P(\phi)}\, \frac{dP}{d\phi} , \qquad g = \frac{2}{d-1} r^2 P(\phi)^2 \,.
\end{equation}
Hence, in these ``fake supergravity'' theories, the asymptotic behavior of the scalar field is determined by the small
$\phi$ behavior of the superpotential.

If we insert the $P_+$ superpotential \eqref{Pplus} into these
equations of motion, we find
\begin{equation}
\phi(r) = \frac{\beta}{r^{d/2}}+\ldots \,,
\end{equation}
so the logarithmic mode is turned off.  As noted above, with this
superpotential the minimum energy result only holds if $\alpha = 0$.

We can instead use one of the generic $P_s$ superpotentials to generate our domain wall, but it was shown in  \cite{Faulkner:2010fh} that any superpotential which is not the critical $P_c$ leads to a naked singularity. Further, the critical superpotential domain wall corresponds to the large $\alpha$ limit of scalar solitons (section \ref{sec:solitons}) and the zero temperature limit of planar black holes with hair (section \ref{sec:poincare}).

When we use the $P_c$ superpotential \eqref{bfsp}, we obtain
the solution
\begin{equation}
\label{fakephi}
 \phi(r) = \frac{\alpha \log r}{r^{d/2}}+\left[\frac{ 8 s_c \sqrt{2(d-1)}+ d( \log(d^2/4)- 1)}{d^2} \alpha -\frac{2}{d} \alpha \log |\alpha| \right] \, \frac{1}{r^{d/2}} + \ldots
\end{equation}
Now the logarithmic mode is present, and $\beta(\alpha)$ takes
the scale invariant form  $\beta(\alpha)  = k \alpha - \frac{2}{d}
\alpha \log \alpha$, which is expected due to the scale invariance of the
equations of motion \eqref{eomfake}.  We also have
\begin{equation}
g = r^2+\frac{d \alpha^2 (\log r)^2}{2 (d-1) r^{d-2}}+\frac{\alpha (d \beta -\alpha) \log r}{(d-1) r^{d-2}}-\frac{M_0}{r^{d-2}}+\ldots \,,
\end{equation}
where
\begin{equation}
\label{M0}
M_0 = -\frac{\left(8 \sqrt{2(d-1)} \, s_c+2 d \log \left(\frac{d}{2}\right) \right) \alpha^2}{d^2 (d-1) }
-\frac{2 \alpha^2 \log \alpha }{d(1-d)}
+\frac{2 \alpha  \beta }{d-1}
-\frac{d \beta^2}{2 (d-1)} \,.
\end{equation}
The energy density for such planar solutions is
\begin{equation}
\frac{E}{V} = \frac{d-1}{2} M_0 -\alpha \beta +\frac{d \beta^2}{4}+W \,,\label{neumannenergy}
\end{equation}
so using \eqref{M0}, we find
\begin{equation}
\frac{E}{V} = W -\frac{4 s_c\sqrt{2(d-1)} + d \log(d/2)}{d^2} \alpha^2 +\frac{1}{d} \alpha^2 \log |\alpha| \,.
\end{equation}
Thus, these solutions saturate the bound \eqref{bfboundfinal}.


\section{Scalar Solitons}
\label{sec:solitons}

To relate the bound \eqref{bfboundfinal} to the conjecture of
\cite{HH2004}, we now analyze the behavior of static, spherical
solitons when $m^2 = m^2_{BF}$.

For an ansatz of the form
\begin{equation}
ds^2=-h(r)e^{-2\chi(r)}dt^2+ \frac{dr^2}{h(r)}+r^2 d\Omega_{d-1}^2 ,
\end{equation}
the equations of motion are
\begin{equation}
h\phi''+\left(\frac{(d-1)h}{r}+\frac{r}{d-1}\phi'^2h+h'\right)\phi'   =  \frac{dV}{d\phi}\,,
\end{equation}
\begin{equation}
(d-2)(1-h)-r h'-\frac{r^2}{d-1}\phi'^2 h =  \frac{2}{d-1}r^2 V,
\end{equation}
\begin{equation}
\chi' = -\frac{1}{d-2} r \phi'^2 .
\end{equation}
Regularity at the origin requires $h(0)= 1$ and $h'(0) =\phi' (0) =\chi'(0)=0$.

For small $\alpha$, one can show analytically that
\begin{equation}
\label{smalla}
\beta_0(\alpha) = -\left(\psi(d/4) +\gamma  \right) \alpha +O(\alpha^3)\,
\end{equation}
where $\gamma$ is Euler's constant and $\psi(z)$ is the digamma function.
Here the slope is  positive for $d=3$,  vanishes in $d=4$,
and is negative for $d \geq 5$.
For linear boundary conditions
$\beta = f \alpha$ (i.e., a double trace deformation), we have
\begin{equation}
\V(\alpha) = \frac{1}{2} \left(f+\psi(d/4) +\gamma  \right)
\alpha^2 \,,
\end{equation}
so in this perturbative regime, stability requires
\begin{equation}
f \geq -\left(\psi(d/4) +\gamma  \right) \,.
\end{equation}
This agrees with the linearized stability analysis of \cite{Ishibashi:2004wx}.

\begin{figure}
\begin{picture}(0,0)
\put(-210,-70){\scriptsize{$\beta_0$}} \put(-102,-130){\scriptsize$\alpha$}
\put(3,-70){\scriptsize{$W_0$}} \put(113,-130){\scriptsize$\alpha$}
\end{picture}
\begin{center}
$\begin{array}{cc}
\includegraphics[width=2.5in]{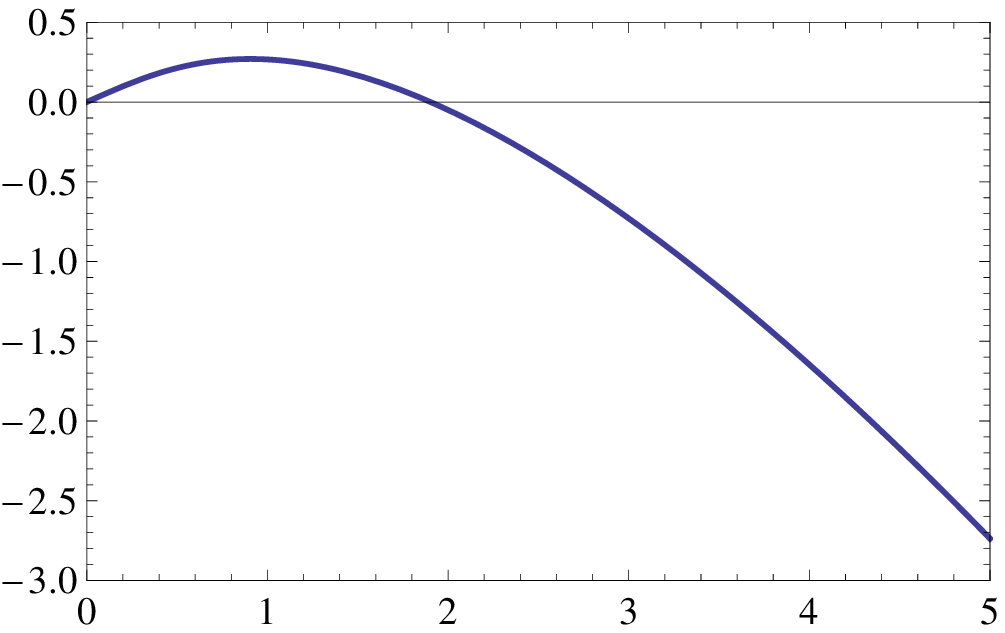}& \hspace{1cm}
\includegraphics[width=2.5in]{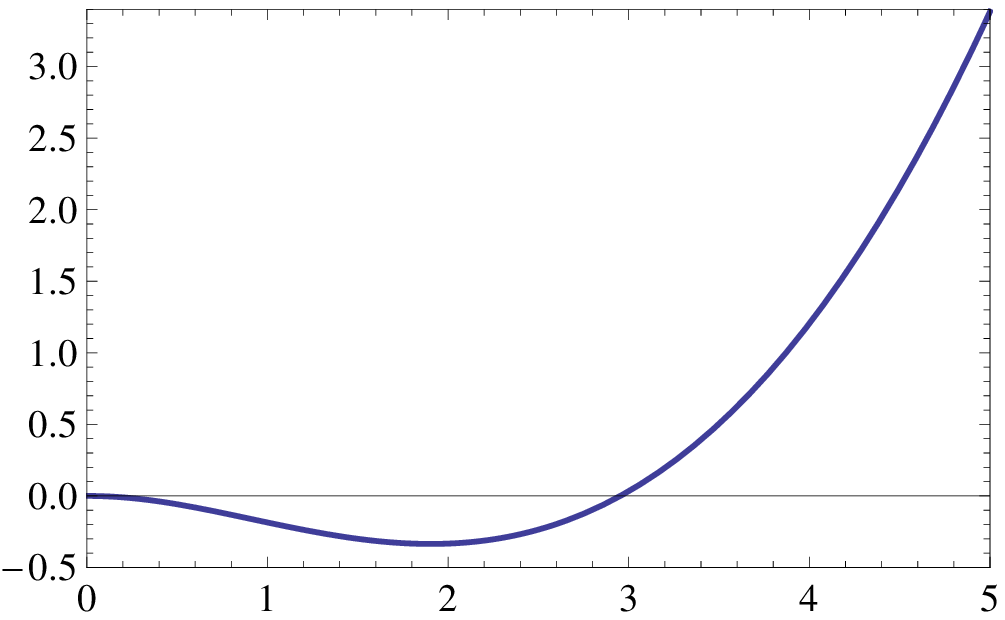}
\end{array}$
\end{center}
\caption{Here we plot the soliton curve $\beta_0(\alpha)$ and the
corresponding $W_0(\alpha)$ in the case \eqref{simpleV}.} \label{soliton}
\end{figure}

For non-perturbative stability, we need the full nonlinear solution,
which can be found numerically.
For example, the soliton curve $\beta_0(\alpha)$ in the simple case
\eqref{simpleV} is plotted in Figure~\ref{soliton}.

Following the arguments of \cite{Faulkner:2010fh}, we expect that the
large $\alpha$ limit turns global solitons into the boost-invariant $P_c$ domain wall of section \ref{fakesugra}.
Thus for large $\alpha$, the soliton curve should take the scale invariant form
\begin{equation}
\beta_0 \sim -\left(2 c_0+\frac{1}{d}\right) \alpha - \frac{2}{d} \alpha \log |\alpha| \,,
\end{equation}
for some constant $c_0$.  This implies
\begin{equation}
\label{W0large}
W_0(\alpha) \sim c_0 \alpha^2 + \frac{1}{d} \alpha^2 \log |\alpha| \,.
\end{equation}
Because of the overall negative sign in the definition \eqref{W0def},
the sign of the logarithmic term is opposite to that of \eqref{bfbc},
so $W_0(\alpha)$ does not take the scale invariant form
(in contrast to $m^2 \neq m^2_{BF}$).  Note however, that the
coefficient of the logarithmic term matches that which appears in
\eqref{bfboundfinal}.  Also, since the logarithmic term
dominates at large $\alpha$, this sign ensures that $W_0$ is bounded from
below.

These expectations about the large $\alpha$ behavior of the solitons are confirmed by our numerical results (see for
example Figure~\ref{soliton}, where $c_0 = -0.45$).  In all examples that we have tested, we find the general relation
\begin{equation}
c_0 = -\frac{4 s_c \sqrt{2(d-1)} + d \log(d/2)}{d^2} \,.
\end{equation}
 It follows from this and \eqref{bfboundfinal} that when $\V = W+W_0$ has a global minimum, the energy is bounded from below.  This proves the conjecture of \cite{HH2004} in the case where the BF
bound is saturated.
\section{Poincar\'e AdS and Finite Temperature}
\label{sec:poincare}

In this section, we examine stability in asymptotically Poincar\'e AdS and further discuss our results from the dual field theory perspective.  We generalize to finite temperature and demonstrate the existence of a phase transition at the critical temperature.

\begin{figure}
\begin{center}
\includegraphics[scale=.735]{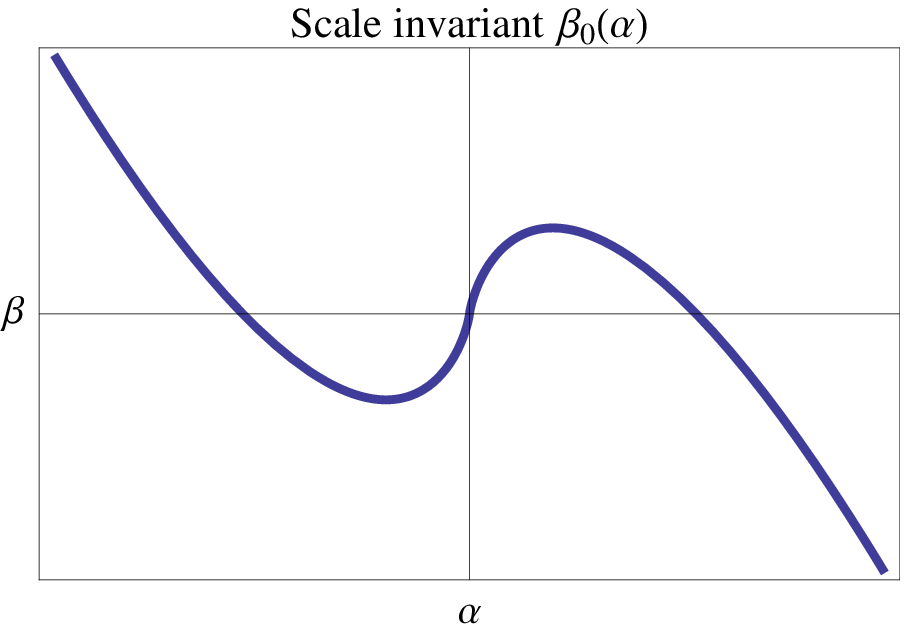} \hspace{1cm}
\includegraphics[scale=.75]{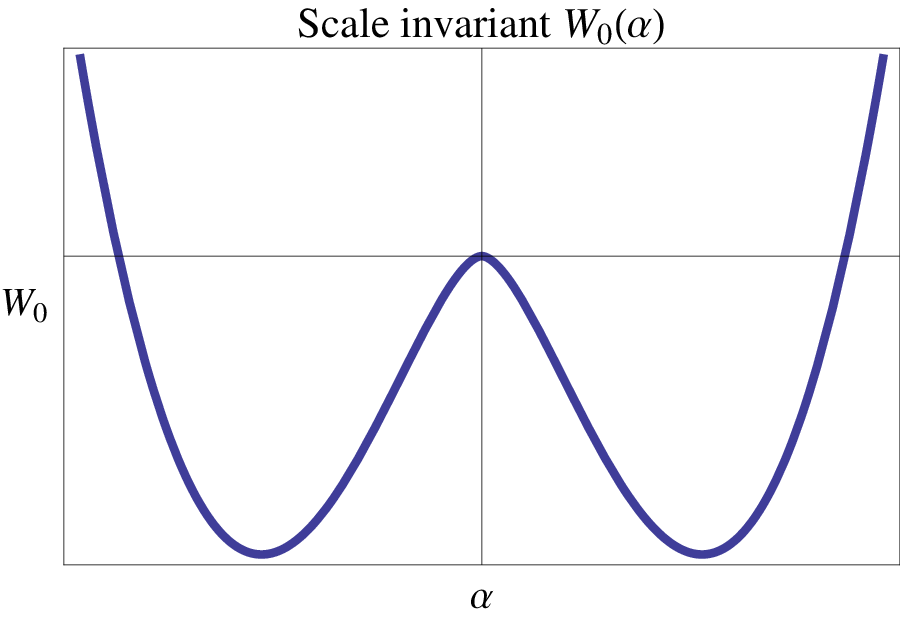}\end{center}
\vspace{-.5cm}
\caption{\label{neumannW0} The left plot shows the planar soliton curve $\beta_0$ (for some generic $k_0$), which always takes the scale invariant form.   The right plot shows the corresponding $W_0$, which does not have the same form as the scale invariant $W(\alpha)$ due to the minus sign in the definition \eqref{W0def}. Note that in the undeformed theory ($W = 0$), the effective potential is precisely $W_0$, and there is always an ordered ground state.}
\end{figure}

Explicitly, we are interested in static, plane-symmetric solutions of the form
\be
ds^2=r^2\eta_{\mu\nu}dx^\mu dx^\nu+\frac{dr^2}{r^2}+\ldots,~\phi = \frac{\alpha \log(r/\Lambda)}{r^{d/2}}+\frac{\beta}{r^{d/2}}+\ldots
\ee
Since we are not in global AdS, there is no longer a natural scale in the theory, and so we will always define the logarithmic terms at a cutoff scale $\Lambda$.  Note that this cutoff scale appears whenever we had a logarithm in global AdS so as to make the argument dimensionless.  For example, the scale transformation acts as
\be
r\rightarrow c r,~ \alpha \rightarrow c^{d/2}\alpha,~\beta \rightarrow c^{d/2}(\beta - \alpha \log c)\,, \label{scaling}
\ee
and is unbroken under the boundary conditions
\be
\beta = k \alpha - \frac{\alpha}{d}\log(\alpha^2/\Lambda^{d}).\label{scalebc}
\ee
In the gauge $g_{ii}=r^2$, the metric behaves asymptotically as
\be
g^{rr}=r^2\left(1 +\frac{\alpha \log(r/\Lambda)(-2\alpha+2 d \beta +d \alpha \log(r/\Lambda))}{2(d-1)r^d} -\frac{M_0}{r^d}+\ldots\right),
\ee
\be
-g_{tt}=r^2\left(1-\frac{2\alpha^2-2d\alpha\beta+d^2\beta^2}{2d(d-1)r^d}-\frac{M_0}{r^d} + \ldots \right).
\ee
The energy of these static solutions with hair is again given by \eqref{neumannenergy}. For zero temperature configurations, we know the solution will be the $P_c$ fake supergravity domain wall, and therefore the planar soliton curve $\beta_0(\alpha)$ takes exactly the scale-invariant form \eqref{scalebc} with some $k=  k_0$ determined by $V(\phi)$.  We can easily integrate this to find
\be
W_0(\alpha)=-\left(\frac{k_0}{2}+\frac{1}{2d}\right)\alpha^2+\frac{\alpha^2\log(\alpha^2/\Lambda^d)}{2d}.
\ee
We plot $\beta_0$ and $W_0$ in Figure \ref{neumannW0}.

 \subsection{Double-Trace Deformations}
For double-trace deformations $\beta = f \alpha$, the coupling has the simple energy scale dependence
\be
f(\mu)=f(\Lambda)+\log(\mu/\Lambda)\,,
\ee
and any nonzero $f_0 \equiv f(\Lambda)$ can be absorbed in a redefinition of the UV scale. This log running is what breaks the conformal symmetries even  for the Neumann theory. There is an IR divergence, and a critical scale $\mu =e^{-f_0}\Lambda \equiv \Lambda_f$ where the coupling changes sign. This is the scale where an instability of the AdS vacuum sets in. We can also see this in the Green's function for $\phi$ in the exact AdS background,
 \be
 G(\omega)=\frac{1}{\log(-i\omega/\Lambda)+f}=\frac{1}{\log(-i\omega/\Lambda_f)}=\frac{i\Lambda_f}{\omega-i\Lambda_f}+\op(1),
 \ee
 which has a pole in the upper half plane.  
 
 The full non-linear effective potential is
 \be
 \V(\alpha)=\frac{-1/d-k_0+f}{2}\alpha^2+\frac{\alpha^2
 }{2d}\log(\alpha^2/\Lambda^d),
 \ee
 which has a global minimum at
 \be
 \alpha_*=\pm \left( e^{k_0-f}\Lambda\right)^{d/2},~\V(\alpha_{*})=-\frac{1}{2d}\left(e^{k_0-f}\Lambda \right)^d
 \ee
 regardless of the value of $k_0$ or $f$!  Thus, for double-trace deformations, the AdS vacuum is always unstable, but there is still a stable ground state.  As usual, the precise nature of this ground state depends on the full nonlinear structure of $V(\phi)$ as it will correspond to a fake supergravity domain wall.

\subsection{Finite Temperature}
\begin{figure}
\begin{center}
\includegraphics[scale=.75]{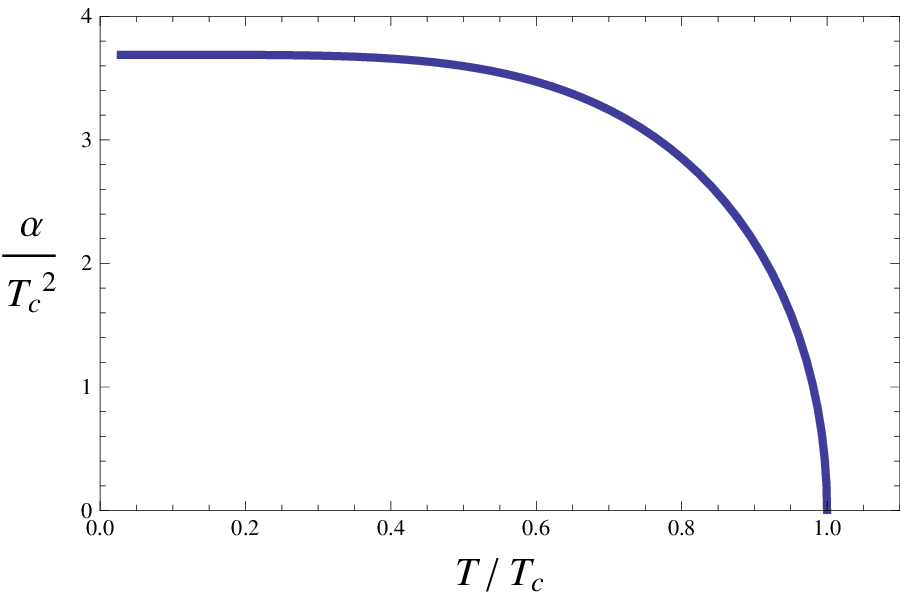} \hspace{1cm}
\includegraphics[scale=.75]{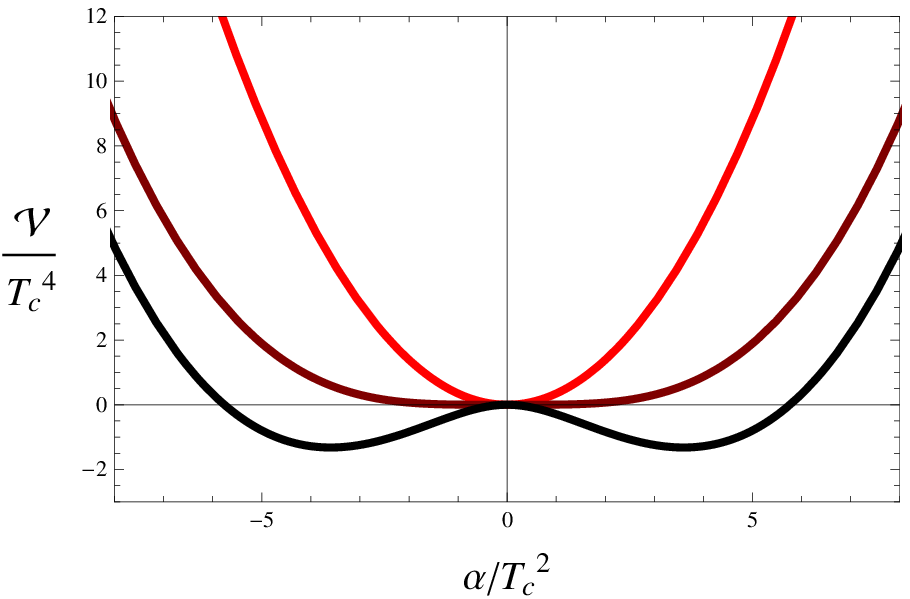}
\end{center}
\vspace{-.5cm}
\caption{\label{atc} The left plot shows the order parameter $\alpha$ as we lower the temperature. Note that $\alpha \propto (T_c-T)^{1/2}$ as expected.  On the right side,  we plot $\V=W_0+W$ at  $T/T_c=(2,~1,~1/2.)$. This is for the theory in  $AdS_5$ with $V(\phi)=-6-2\phi^2+\phi^4$.}
\end{figure}

The instability of the AdS vacuum described in the previous section occurs at zero temperature, and it persists for low temperature.  However, heating the system up enough will lift the instability.  As in \cite{Faulkner:2010gj}, we can identify the critical temperature by looking for a static normalizeable mode for the scalar field in the background of AdS-Schwarzschild. This locates the temperature at which the zero-momentum quasinormal mode moves from the upper to the lower complex plane, which is precisely $T_c$. The static linearized wave equation for $\delta\phi = \phi(r)$ on the AdS-Schwarzschild background is
\be
\phi''+\left(\frac{1}{r}+\frac{d r}{g}\right)\phi'+\frac{d^2}{4g}\phi=0,~g=r^2\left(1-\left(\frac{4\pi T}{dr} \right)^d\right).
\ee
The solution which is smooth on the horizon is
\be
\phi_c=\frac{2 \alpha}{dr^{d/2}}Q_{-1/2}(1-2(4\pi T/d r)^d),
\ee
where $Q_\nu(z)$ is the Legendre function of the second kind. This behaves at large $r$ as
\be
\phi_c=\frac{\alpha \log(r/\Lambda)}{r^{d/2}}-\frac{\alpha\log\left(4^{1-2/d}\pi T/d\Lambda\right)}{r^{d/2}}+\ldots\,,
\ee
which implies that the critical temperature is
\be
\label{Tc}
T_c=\frac{4^{2/d-1}d}{\pi}e^{-f}\Lambda\propto \Lambda_f \,.
\ee
This is the location of a second order phase transition.  We can calculate the behavior of the order parameter and the full off-shell potential by constructing numerical solutions (see Figure \ref{atc}). The system behaves much like the case away from the BF bound studied in \cite{Faulkner:2010gj}, because the system near $T_c$ is governed by the temperature and not the order parameter.
We confirm by calculating $\V$ that the second order phase transition is not masked by a first order transition.

\section{Deforming the Dirichlet theory}
\label{sec:dirichlet}

The previous sections have studied deformations of the Neumann theory $\beta = 0$.  In this section, we discuss some analogous results for deformations of the Dirichlet theory $\alpha = 0$.

In \eqref{dW}, we chose a scalar boundary condition $\beta = \beta(\alpha)$, but it is equally valid to take instead $\alpha = \alpha(\beta)$, which we again express as
\begin{equation}
\alpha = \frac{dW}{d\beta}
\end{equation}
for some function $W(\beta)$.  This boundary condition corresponds to a deformation of the Dirichlet theory by a term $W(\op)$, where the operator $\op$ has conformal dimension $\Delta = \lambda_+$ (or $\Delta = d/2$ at the BF bound).

Denoting the soliton curve as $\alpha_0(\beta)$, the effective potential is
\be
\V(\beta)=W_0(\beta)-W(\beta),~W_0(\beta)=+\int_0^\beta \alpha_0(\tilde \beta)d\tilde\beta \,.\ee
Once again, this satisfies the properties that extrema of $\V$ give solutions satisfying our boundary conditions and the value of $\V$ at the extremum is the energy of the corresponding soliton.

Focusing on the plane-symmetric solutions, the energy is
\be
\frac{E}{V}=\frac{d-1}{2}M_0+\frac{d}{4}\beta^2-W(\beta) \,,
\ee
which satisfies a bound analogous to \eqref{bfboundfinal},
\begin{equation}
\label{dirichletbound}
E \geq \oint\left[\alpha \beta-W(\beta) -\frac{4 s\sqrt{2(d-1)} + d \log(d/2)}{d^2}
\alpha^2 +\frac{1}{2d} \alpha^2 \log\alpha^2 \right] \,.
\end{equation}
Note that since there is no term on the right hand side of \eqref{dirichletbound} corresponding to $W_0(\beta)$, the spinor charge calculation does not seem to produce an energy bound in terms of $\V(\beta)$\footnote{This point is not related to the fact that the BF bound is saturated.  A similar statement would hold for other masses in the range \eqref{range}.}.  Nevertheless, in what follows we shall assume that the designer gravity  conjecture holds, so that stability of the theory is still determined by the effective potential $\V$.

\begin{figure}
\begin{picture}(0,0)
\put(-80,-42){$ \scriptstyle (0,\,-1)$}
\put(-95,-83){$ \scriptstyle (-1,\,-1)$}
\put(-150,-75){$ \scriptstyle (-1,\,1)$}
\put(-150,-118){$ \scriptstyle (0,\,1)$}
\put(72,-105){$ \scriptstyle (0,\,-1)$}
\put(128,-30){$ \scriptstyle (-1,\,-1)$}
\put(85,-31){$ \scriptstyle (-1,\,1)$}
\put(144,-105){$ \scriptstyle (0,\,1)$}
\end{picture}
\begin{center}
\includegraphics[scale=.735]{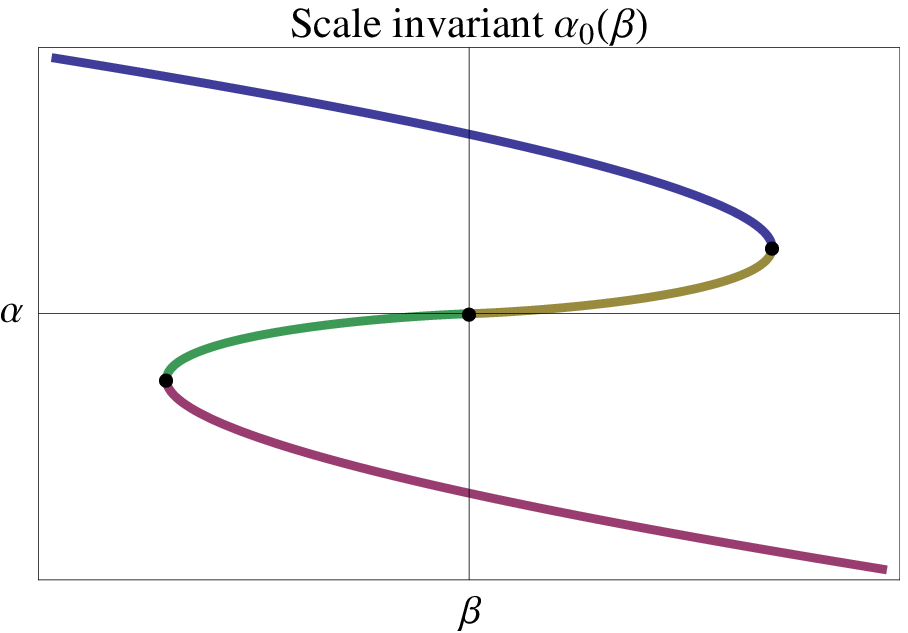} \hspace{1cm}
\includegraphics[scale=.75]{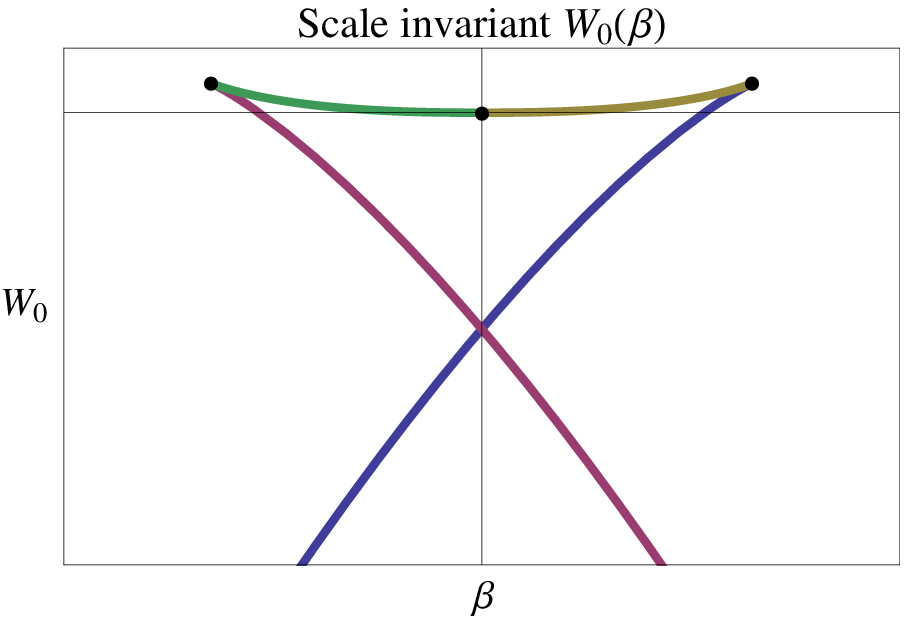}\end{center}
\vspace{-.5cm}
\caption{\label{branchstr} These plots show the four branches of $\alpha_0(\beta)$ and $W_0(\beta)$.  The curves are labeled by the corresponding values of $(n,\sigma)$.}
\end{figure}

To calculate $W_0(\beta)$ we must invert (\ref{scalebc}), which is not one-to-one. We therefore find the solution in four branches,
\be
\alpha_0(\beta) = -\frac{d\beta}{2 w_n(z) },~z=\sigma\frac{\beta d}{2 (e^{k_0} \Lambda)^{d/2}} ,~n=(-1,0),~\sigma=\pm 1
\ee
which then implies
\be
W_0(\beta)=-\frac{d\beta^2}{8 w_n(z)^2}\left(1+2w_n(z) \right) \,.
\ee
Here $w_n(z)$ is the generalized Lambert function, defined as the $n^\mathrm{th}$ solution to $z=w e^w$. These four branches are shown in Figure \ref{branchstr}.
The $n=0$ branches follow $\beta \rightarrow \infty$, so using the fact that
\begin{equation}
w_0(z) = \log z-\log(\log z)+O\left(\frac{\log(\log z)}{\log z}\right)\,,
\end{equation}
we find
\be
W_0(\beta) = - \frac{d \beta^2}{2 \log(\beta^2/\Lambda^d)}+\ldots\label{w0dir}
\ee  Note that even though $W_0$ is unbounded from below, this does \emph{not} imply that the $\alpha=0$ theory is unstable, as in that case all we require is the existence of a $P_+$ to prove positivity of the energy.

Let us now consider double-trace deformations of the Dirichlet theory\footnote{Note that our definition of $\alpha$ differs by a sign from that of \cite{witten}, but our definition of $f$ is  the same.}, $\alpha=-f \beta$, corresponding to $\Delta S_{CFT}= - \frac{f}{2}\int\op^2$. As pointed out in \cite{witten}, $f$ runs as
\be
f(\mu)=\frac{f(\Lambda)}{1-f(\Lambda)\log(\mu/\Lambda)}
\ee
and  is marginally irrelevant (relevant)  for $f>0$  ($f<0$),  running to a Landau pole at the scale $\mu = \Lambda e^{1/f_0}$ which is above (below) the UV scale $\Lambda$. As shown in \cite{Iqbal:2011aj}, the Green's function for $\phi$ on an AdS background in the deformed theory is
\be
G(\omega)=\frac{1}{-1/\log(-i\omega/\Lambda)+f} \,,
\ee
which has a pole at $\omega=i\Lambda e^{1/f}$, a width corresponding to the Landau pole scale. As we turn on a positive $f$, the pole comes in from $+i\infty$, confirming that it is a marginally irrelevant deformation.  Turning on a negative $f$ brings a pole out of the origin, implying an infrared instability from the marginal deformation.

\begin{figure}
\begin{center}
\includegraphics[scale=1]{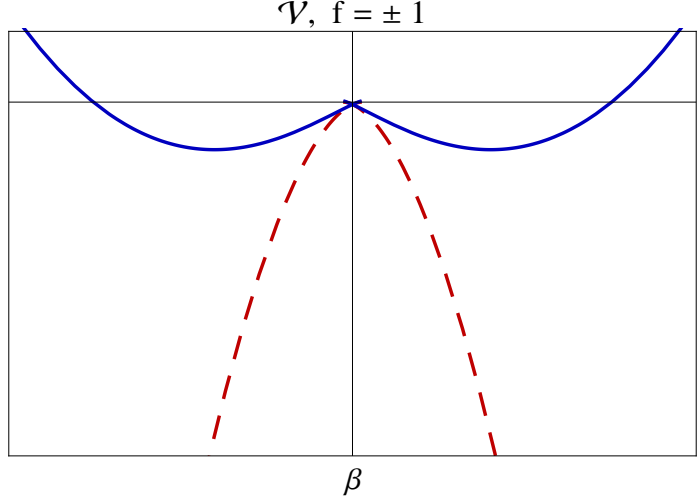}\end{center}
\vspace{-.5cm}
\caption{\label{atc2} The effective potential $\V$ for a double-trace deformation of the Dirichlet theory. The blue curve corresponds to $f>0$ and is bounded from below.  The dashed red curve corresponds to $f<0$ and has no lower bound. }
\end{figure}

To  study the non-perturbative stability of the theory, we look at the effective potential $\V(\beta)$.  Using our result above for $W_0$ in the large $\beta$ limit, we have
\be
\V=W_0-W =+ \frac{f}{2}\beta^2 - \frac{d \beta^2}{4 \log(\beta^2/\Lambda^d)}+\ldots
\ee
and therefore for positive $f$ there is a global minimum at
\be
\beta_*=\pm\frac{( e^{k_0+1/f}\Lambda)^{d/2}}{f}\,,~\V(\beta_*)=-\frac{(e^{k_0+1/f}\Lambda)^d}{2d}\,.
\ee
This is a scale well above our UV scale $\Lambda$, as expected from an irrelevant deformation. For negative $f$, the effective potential $\V$ has \emph{no} global minimum. This suggests that the theory is unstable and has no minimum energy state. This is demonstrated in a plot of $\V$ for both signs of $f$ in Figure
\ref{atc2}.

  This implies that, unless we turn on higher order multitrace terms, there is no endpoint to the IR instability.  This is what we should expect, as we have turned on a negative sign deformation which is not stabilized by any terms in the undeformed effective potential $W_0$. We suspect that the resulting dynamics will be similar to what was found in \cite{Hertog:2005hu},   where a destabilizing marginal perturbation lead to a big crunch in the bulk. It was important in constructing these explicit time-dependent big crunch solutions that the boundary conditions preserved the full conformal symmetry. In our system, where the BF bound is saturated, the linear boundary conditions $\alpha = - f \beta$ are not scale-invariant, but we still expect that there will be a big crunch. Indeed, whenever $\V$ is unbounded from below, we suspect that the system will have a big crunch instability.


\section{Discussion}
\label{sec:discussion}
In this work, we have proved a minimum energy theorem for asymptotically AdS spacetimes containing scalar fields that saturate the Breitenlohner-Freedman bound.  The main result, \eqref{bfboundfinal}, was derived by showing that an appropriately defined spinor charge is positive, and then relating this spinor charge to the physical energy.  The previously troublesome issue of divergent terms in this calculation was resolved by finding a new branch of superpotential solutions \eqref{bfsp}, whose existence is required for the bound to hold.  We then showed that our result further implies that the theory is stable when the full effective potential $\V$ has a global minimum.  This verified the conjecture by the authors of \cite{HH2004} in the case where the scalars saturate the BF bound.

For asymptotically Poincar\'e AdS solutions, the Neumann theory with double trace deformations is always stable, though AdS itself is not the ground state. The ground state is always a domain wall with a nonzero vacuum expectation value of the dual operator $\op$.  At sufficiently high temperature, however, the theory returns to the symmetry-preserving state, and thus there is a phase transition at the critical temperature \eqref{Tc}.  We also explained the endpoint of the instability due to a positive double-trace deformation in the Dirichlet theory. We conjecture that the theory with a negative double-trace deformation is unstable, in that it has no minimum energy state.

Witten's original spinorial proof of the positive energy theorem was motivated by the idea that any supersymmetric theory must be stable,  since then the Hamiltonian can then be expressed as a square of the supercharge.    It is important to note, however, that supersymmetry is not necessary for the derivation of the energy bound \eqref{bfboundfinal}.  The superpotential used to construct the spinor charge does not have to be the ``actual'' superpotential that appears in supersymmetry transformations; we only require that $P(\phi)$ is a real, global solution to \eqref{vtop}.  In any case, the superpotential \eqref{bfsp} is non-analytic and thus would not arise in a supersymmetric theory.  Furthermore, the boundary conditions that we consider do not preserve supersymmetry, and in particular, there are \emph{no} supersymmetric multi-trace deformations when the BF bound is saturated  \cite{susymulti}.

\begin{figure}
\begin{center}
\includegraphics[scale=1]{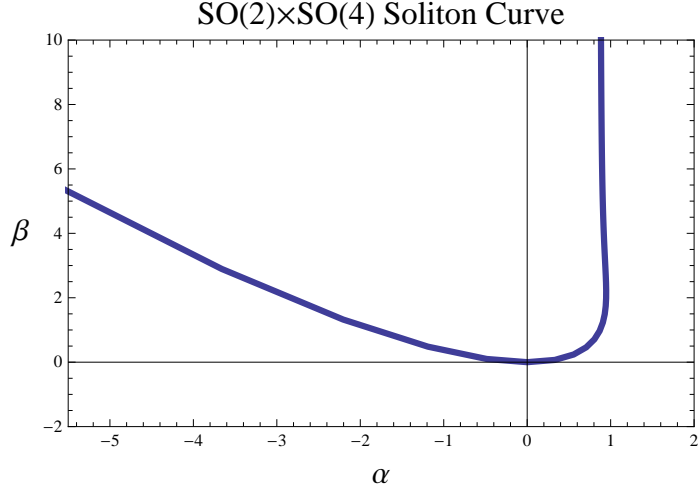}\end{center}
\vspace{-.5cm}
\caption{\label{so2so4curve} Here we plot the ($\alpha,\beta)$ curve for the $SO(2)\times SO(4)$ consistent truncation of \cite{Freedman:1999gk}, which has $V(\phi)=-2 e^{-\phi/\sqrt{3}}\left(2+e^{\sqrt{3}\phi} \right)$. The vertical asymptotics in the upper right corresponds to $\phi(0)\rightarrow +\infty$ and the soliton becoming the 1/2 BPS domain wall while the log-power law behavior on the left  corresponds to $\phi(0)\rightarrow -\infty$ and the soliton becoming a fake supergravity domain wall.}
\end{figure}

 In some of the consistent supersymmetric truncations which include scalar fields saturating the BF bound, the soliton curve is not single-valued as a function of $\beta$ or $\alpha$. In the known examples of consistent $AdS_5 \times S^5$  truncations, we find that the spherical soliton solution sometimes tends towards $\alpha = \mathrm{const.}$ as $\beta \to \infty$.  In the planar limit, this becomes a domain wall solution with $\alpha = 0,~ \beta = \mathrm{const}$.  These domain walls are actually 1/2 BPS and the ten-dimensional description involves smearing the D3 branes in the transverse directions  \cite{Freedman:1999gk}.  These domain walls are honest supergravity domain walls, in that the $P=P_+$ which generates their solution in the manner of (\ref{eomfake}) is the analytic superpotential in the bulk supersymmetry algebra. However, we find that in cases without even potentials, the soliton with $\phi(r=0) \rightarrow +\infty$ approaches the 1/2 BPS domain wall, but the solution with $\phi(r=0) \rightarrow -\infty$ approaches a fake supergravity solution corresponding to a $P_c$ with $s_c\neq 0$. As an example, we present the nontrivial $(\alpha,\beta)$ curve for the $SO(2)\times SO(4)$ scalar of  \cite{Freedman:1999gk} in figure \ref{so2so4curve}. This means that despite not being able to construct a simple $W_0(\alpha)$, we can still find a critical superpotential (by gluing together the critical superpotential for $\phi<0$ and the analytic superpotential for $\phi<0$) and therefore prove an explicit energy bound. It would be very interesting to study the implications of this for deformations of the $\mathcal{N}=4$ theory.

The stability conjecture of \cite{HH2004} in fact contained a second part:  the soliton associated with the global minimum of $\V$ \emph{is} the minimum energy solution.  Note that this does not automatically follow from \eqref{sbound} or \eqref{bfboundfinal}, since the terms on the right hand side of the inequality only approach $\V$ for large $\alpha$.  Hence, proof that the ground state is the minimum energy soliton is still an open issue (it has been shown in \cite{Hollands:2005wt} that the minimum energy solution must be static).  As mentioned above, there are additional cases where the scalar field has a logarithmic branch near the AdS boundary, and so it might be worthwhile to study stability for these other so-called ``resonant'' theories (stability in the case $n =\lambda_+/\lambda_- = 3$ was partially addressed in \cite{Amsel:2006uf}).  It would also be interesting to further understand energy bounds in the Dirichlet theory, as in this case the result from the spinor charge calculation \eqref{dirichletbound} did not lead to an expression related to the effective potential $\V(\beta)$. Finally, it is important to point out that $\Delta \Gamma = W$ is only true at leading order in $1/N$, and it would be interesting to understand how non-planar corrections modify this story.


\section*{Acknowledgments}
It is a pleasure to thank Gary Horowitz, Donald Marolf, and Massimo Porrati for useful discussions. M.M.R.~is supported by the Simons Postdoctoral Fellowship Program. Much of this work was completed while A.A. was at the University of Alberta, with support from the Killam Trust and NSERC.

\appendix

\section{Approaching the BF Bound}
\label{limit}

Another method to obtain stability results in the case $m^2 =
m^2_{BF}$ is to start with the known results of \cite{Faulkner:2010fh} for $m^2 \neq m^2_{BF}$
and take the limit as the mass approaches the BF
bound.
In this appendix it will be useful to distinguish the two cases using
the ``\,\^{}\,'' notation
\begin{equation}
\phi = \frac{\alpha}{r^{\lambda_{-}}} + \frac{\beta}{r^{\lambda_{+}}}
+ \dots \,, \qquad m^2 \neq m^{2}_{BF}
\end{equation}
\begin{equation}
\phi = \frac{\halpha \,\log r}{r^{d/2}} + \frac{\hbeta}{r^{d/2}}
+ \dots \,, \hspace{.4cm} m^2 = m^{2}_{BF} \,.
\end{equation}
Let us define $2 \heps = (\lambda_+-\lambda_-)$ and consider the limit
$\heps \to 0$.  In \cite{Amsel:2006uf} it was noted that the
identification
\begin{equation}
\label{bftrans}
\alpha \to \frac{\hat{\alpha}}{2 \heps}, \quad \beta \to \hat{\beta} -
\frac{\hat{\alpha}}{2\heps},
\quad 2 \heps W \to \hat{W} - \frac{\hat{\alpha}^2}{4\heps} \, .
\end{equation}
correctly transformed the expression for the conserved energy $E
\to\hat E$.  Applying this transformation to the expression for the spinor
charge in \cite{Amsel:2006uf} leads to a term divergent term of the
form $\halpha^2/\heps$.  So this does not yield an energy bound, in
agreement with the statement in section \ref{reviewBF}.

We can now try instead to take the limit of the modified expression for the
energy bound
\begin{equation}
E \geq (\lambda_+ - \lambda_-)\oint \left[W+\frac{\lambda_-}{d} \, s |\alpha|^{d/\lambda_-}\right]
\end{equation}
Applying the transformation \eqref{bftrans} to this expression and
reparametrizing
\begin{equation}
\label{news}
s = 1+\frac{4}{d} \heps \log(2\heps) +\frac{2(d-8 \hat s\sqrt{2(d-1)}
-2 d \log(d/2))}{d^2} \,\heps
\end{equation}
for some constant $\hat s$, the energy bound takes the form
\begin{equation}
\label{bfebound}
\hat E \geq \oint \left[\hat W  -\frac{4 \hat s\sqrt{2(d-1)} + d \log(d/2)}{d^2} \halpha^2+\frac{\halpha^2}{d} \log \halpha \right] \,.
\end{equation}
The $O(1)$ term in \eqref{news} cancels the original $\heps^{-1}$
divergence, but introduces a new term diverging as $\log \heps$.  This
logarithmic divergence is canceled by the $O(\heps \log \heps)$ term
in \eqref{news}, while the $O(\heps)$ term gives a finite
contribution.  Dropping the hat notation, this exactly reproduces the result
\eqref{bfboundfinal}.

We can also attempt to take the limit of the generalized
superpotential from \cite{Faulkner:2010fh},
\begin{equation}
P_s(\phi) = \sqrt{\frac{d-1}{2}} +\frac{\lambda_-}{2\sqrt{2(d-1)}} \,
\phi^2-\frac{\lambda_- (\lambda_+ -\lambda_-)}{d \sqrt{2(d-1)}} \,s |\phi|^{d/\lambda_-}+\ldots \,.
\end{equation}
Since
\begin{equation}
\phi^{d/\lambda_-} = \phi^2 \left(1+ \frac{4\heps}{d} \log
\phi+O(\heps^2) \right)
\end{equation}
a reasonable guess \cite{Faulkner:2010fh} is to consider solutions of the form
$P = \sqrt{(d-1)/2}+p_0 \phi^2+p_1 \phi^2 \log \phi +\ldots$.  However,
it is straightforward to check that expansions of this form do not
represent a well defined series solution near $\phi = 0$ (unless $p_1 = 0$).
Furthermore, as noted in the text above, a new correction of the form
$\phi^2 \log \phi$ is not of the right form to cancel the divergence
that appeared in the spinor charge calculation of \cite{Amsel:2006uf}.
Thus, taking the $\epsilon \to 0$ limit is a bit more subtle.

We begin by considering masses near the BF bound,
with $\heps \ll 1$ but non-zero.
Solutions to \eqref{vtop} then have the general pattern
\begin{eqnarray}
\label{spsum}
P(\phi) &=& \sqrt{\frac{d-1}{2}}+\frac{d/2-\heps}{2\sqrt{2(d-1)}} \phi^2+ \sum_{n = 0}^N c_n \phi^{\frac{d + 2 n \heps}{d/2 -  \heps}}+O(\phi^4) \,,
\end{eqnarray}
where
\begin{equation}
c_0 = -\frac{\heps (d-2 \heps)}{d \sqrt{2(d-1)}} \, s
\end{equation}
and $N = \left[\frac{d - 4 \heps}{2 \heps}\right]$.
Note that
\begin{equation}
\lim_{\heps \to 0} \frac{d + 2 n \heps}{d/2 -  \heps} = 2 \,, \qquad \lim_{\heps \to 0} N = \infty \,.
\end{equation}
So there are an infinite number of such correction terms which converge
to $O(\phi^2)$ as we approach the BF bound.

The remaining coefficients $c_n$ are of course fixed by solving \eqref{vtop}.  The result for the first three are
\begin{eqnarray}
c_1 &=& -\frac{\heps s^2}{\sqrt{2(d-1)}}  \\
c_2 &=& -\frac{\heps (d+2 \heps) s^3}{\sqrt{2(d-1)}(d-2 \heps )}  \\
c_3 &=& -\frac{\heps(d+2 \heps) (3d+10 \heps)s^4}{3 \sqrt{2(d-1)}(d-2
\heps )^2}
\end{eqnarray}
Now, based on the results of the previous section (and in particular
\eqref{news}) it will be useful to reparametrize
\begin{equation}
s = 1+\frac{4}{d} \heps \log(2\heps)+\frac{2}{d}\,(1+2 d \bar s) \heps
\end{equation}
for some arbitrary parameter $\bar s$.  Substituting this into the $c_n$ given above, expanding for small $\heps$, and collecting terms, we find
\begin{eqnarray}
P(\phi) &=& 1+\frac{d/2-\heps}{2\sqrt{2(d-1)}} \phi^2+
\phi^{\frac{d}{d/2 -\heps}} \left[-\frac{\heps}{\sqrt{2(d-1)}}(1+\zeta+\zeta^2+\zeta^3+\ldots)\right. \nonumber \\
&&-\frac{2 \sqrt{2}\heps^2 \log(2\heps)}{d \sqrt{d-1}}(1+2\zeta+3\zeta^2+4\zeta^3+\ldots)
-\frac{2 \sqrt{2} \heps^2 \bar s}{\sqrt{d-1}} (1+2\zeta+3\zeta^2+4\zeta^3+\ldots)\nonumber \\
&&\left.+\frac{2 \sqrt{2}\heps^2}{d \sqrt{d-1}}\left(-\zeta-\frac{5}{2}\zeta^2-\frac{13}{3} \zeta^3+\ldots\right) + \ldots
\right]+O(\phi^4)\,,
\end{eqnarray}
where $\zeta \equiv \phi^{\frac{2  \heps}{d/2 - \heps}}$.
Treating $\phi$ as small, so that $\zeta \ll 1$,
we sum the series in $\zeta$ to get
\begin{eqnarray}
P(\phi) &=& 1+\frac{d/2-\heps}{2\sqrt{2(d-1)}} \phi^2+
\phi^{\frac{d}{d/2 -\heps}}
\left[-\frac{\heps}{\sqrt{2(d-1)}}\frac{1}{1-\zeta}
+\frac{2 \sqrt{2}\heps^2}{d \sqrt{d-1}}
\frac{\log(1-\zeta)-\log(2\heps)}{(1-\zeta)^2} \right.\nonumber \\
&& \hspace{5.5cm}\left.-\frac{2 \sqrt{2} \heps^2 \bar s}{\sqrt{d-1}}\frac{1}{(1-\zeta)^2}\right]+\ldots \end{eqnarray}
At this point it is safe to take $\heps \to 0$, with the result
\begin{equation}
\label{splim}
P(\phi) = 1+p_0 \phi^2+p_0
\frac{\phi^2}{\log \phi}
+p_0 \frac{\phi^2 \log|\log\phi|}{(\log
\phi)^2}+ p_0 (\log(2/d)-\bar s d) \frac{\phi^2}{(\log
\phi)^2}+\ldots
\end{equation}
where $p_0$ is given in \eqref{p0}.  Finally, rewriting
\begin{equation}
\bar s = \frac{d \log(2/d) - 4 \hs \sqrt{2(d-1)}}{d^2}
\end{equation}
reproduces \eqref{bfsp}.

\end{document}